\documentclass[prl,twocolumn]{revtex4}
\usepackage{amsfonts}
\usepackage{amsmath}
\usepackage{amssymb}
\usepackage{graphicx}

\newcommand{\xh}{\mathbf{x}}
\newcommand{\yh}{\mathbf{y}}
\newcommand{\rv}{\mathbf{r}}
\newcommand{\Qv}{\mathbf{Q}}
\newcommand{\Tr}{{\rm Tr}}
\newcommand{\R}{\rangle}
\renewcommand{\L}{\langle}

\newcommand{\dxx}{$d_{x^2-y^2}$}
\newcommand{\dxy}{$d_{xy}$}

\begin{document}
\title{Antiferromagnetism and superconductivity in layered organic conductors: Variational cluster approach}
\author{P.~Sahebsara and D.~S\'en\'echal}
\affiliation{D\'{e}partement de physique and Regroupement qu\'{e}b\'{e}cois sur les mat\'{e}riaux de pointe, Universit\'{e} de Sherbrooke, Sherbrooke, Qu\'{e}bec, Canada, J1K 2R1}
\date{April 2006}

\begin{abstract}
The $\kappa$-(ET)$_2$X layered conductors (where ET stands for BEDT-TTF) are studied within the dimer model as a function of the diagonal hopping $t^\prime$ and Hubbard repulsion $U$.
Antiferromagnetism and $d$-wave superconductivity are investigated at zero temperature using variational cluster perturbation theory (V-CPT).
For large $U$, N\'eel antiferromagnetism exists for $t' < t'_{c2}$, with $t'_{c2}\sim 0.9$.
For fixed $t'$, as $U$ is decreased (or pressure increased), a $d_{x^2-y^2}$ superconducting phase appears. 
When $U$ is decreased further, the a $d_{xy}$ order takes over.
There is a critical value of $t'_{c1}\sim 0.8$ of $t'$ beyond which the AF and dSC phases are separated by Mott disordered phase.
\end{abstract} 

\maketitle

The proximity of antiferromagnetism (AF) and d-wave superconductivity (dSC) is a central and universal feature of high-temperature superconductors, and leads naturally to the hypothesis that the mechanisms behind the two phases are intimately related.
This proximity is also observed in the layered organic conductor $\kappa$-(ET)$_2$\-Cu\-[N(CN)$_2$]\-Cl, an antiferromagnet that transits to a superconducting phase upon applying pressure\cite{Lefebvre:2000} (here ET stands for BEDT-TTF).
Other compounds of the same family, $\kappa$-(ET)$_2$\-Cu\-(NCS)$_2$ and $\kappa$-(ET)$_2$\-Cu\-[N(CN)$_2$]\-Br, are superconductors with a critical temperature near 10K at ambient pressure.
However, another member of this family, $\kappa$-(ET)$_2$\-Cu$_2$\-(CN)$_3$, displays no sign of AF order, but becomes superconducting upon applying pressure\cite{Shimizu:2003, Kurosaki:2005}.
The character of the superconductivity in these compounds is still controversial.
While many experiments indicate that the SC gap has nodes (presumably $d$-wave), others are interpreted as favoring a nodeless gap.
The literature on the subject is rich, and we refer to a recent review article\cite{Powell:2006} for references.

\begin{figure}[tbp]
\centerline{\includegraphics[width=8.5cm]{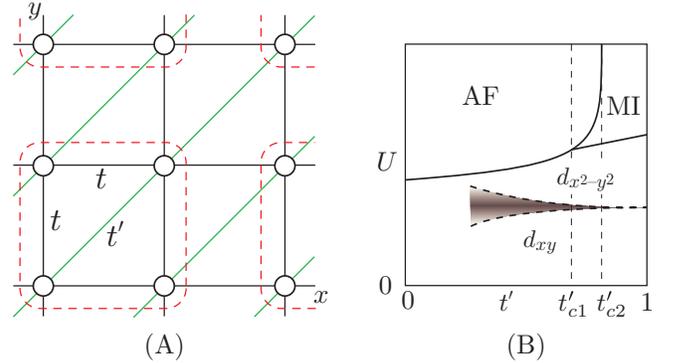}}
\caption{(Color online) (A) : Schematic view of the hopping terms in the dimer model. V-CPT uses a tiling of clusters such as the 4-site cluster drawn here (dashed lines). (B) : Qualitative phase diagram inferred from our calculations (not to scale). See text for details.
}
\label{fig_dimer}
\vglue-4mm
\end{figure}

The interplay of AF and dSC orders, common to both high-$T_c$ and $\kappa$-ET materials, cannot be fortuitous and must be a robust feature that can be captured in a simple model of these strongly correlated systems.
$\kappa$-ET compounds consist of orthogonally aligned ET dimers that form conducting layers sandwiched between insulating polymerized anion layers.
The simplest theoretical description of these complex compounds is the so-called dimer Hubbard model\cite{Kino:1996, McKenzie:1998} (Fig.~\ref{fig_dimer}A) in which a single bonding orbital is considered on each dimer, occupied by one electron on average, with the Hamiltonian
\begin{eqnarray}
H=t\sum_{\langle \rv\rv'\rangle,\sigma}c^\dagger_{\rv\sigma}c_{\rv'\sigma} +
 t' \sum_{[\rv\rv'],\sigma} c^\dagger_{\rv\sigma}c_{\rv'\sigma} + U \sum_\rv n_{\rv\uparrow} n_{\rv\downarrow}
\end{eqnarray} 
where $c_{\rv\sigma}$ ($c^\dagger_{\rv\sigma}$) creates an electron (hole) at dimer site $\rv$ on a square lattice with spin projection $\sigma$, and $n_{\rv\sigma}$=$c^\dagger_{\rv\sigma}c_{\rv\sigma}$ is the hole number operator. $\langle \rv\rv' \rangle$ ($[\rv\rv']$) indicates nearest- (next-nearest)-neighbor bonds.
As the ratio $t'/t$ grows from 0 towards 1, N\'eel AF is increasingly frustrated.
The values of $t'$ for the Br, Cl, NCS and CN$_3$ compounds  are thought to be roughly $0.5-0.65$, $0.75$, $0.75-0.85$ and 1 respectively~\cite{McKenzie:1998,Shimizu:2003}.
The local Coulomb repulsion $U$ is hard to calculate from first principles, but is estimated to be of the same order of magnitude as the band-width\cite{Powell:2006}.
We assume that applying pressure has the effect of increasing $t$ and $t'$ in proportion, without affecting $U$; thus, for fixed $t$, we refer to $t/U$ as the pressure\cite{Kino:1996, McKenzie:1998}.
Different anions correspond also to different chemical pressures, so that, at ambient pressure, the various members of the family of compounds correspond to different points on the $(U,t')$ plane.
Henceforth we will set $t$ to 1, thereby measuring $t'$ and $U$ in units of $t$.

The dimer model has been studied with a variety of methods.
For instance, The fluctuation exchange approximation (FLEX)\cite{flex_REF} predicts the existence of d-wave superconductivity based on diverging susceptibilities.
Quantum Monte Carlo calculations\cite{Kuroki:1999} find enhanced dSC correlations.
Ground state wave-functions obtained from variational methods\cite{variational_GS} predict a transition between a dSC and a spin liquid phase, akin to a Mott transition, in the highly frustrated case.
The spin liquid phase was also predicted by variational methods, and through a gauge theory of the Hubbard model\cite{Motrunich:2005}.
Variational calculations based on the dimer model plus additional exchange terms reveal a transition between dSC and an AF insulator\cite{RVB}.
Cellular Dynamical Mean Field Theory (CDMFT)\cite{Parcollet:2004} has also been used to reveal the Mott transition in that case, without treating the broken symmetry phases.

We report results of a zero-temperature study of the dimer model using Variational Cluster Perturbation Theory (V-CPT)\cite{Potthoff:2003}.
This method captures short-range correlations exactly, treats broken symmetries with a rigorous dynamical variational principle, and provides dynamical information (the spectral function).
It has been recently applied with success to the broken symmetry phases in models of high-$T_c$ superconductors\cite{Senechal:2005, Aichhorn:2005}: transitions between AF and dSC states were found as a function of doping, both for hole- and electron-doped materials, at roughly the correct doping levels.
Let us summarize our results (see Fig.~\ref{fig_dimer}B): Beyond the normal (disordered) state, the following states have been investigated: an AF (N\'eel) state, a dSC state with \dxx\ symmetry and an extended $s$-wave state which, for reasons explained below, we will call \dxy.
For all values of $t'$ we find a \dxx\ phase at low $U$.
For $t'$ smaller than a critical value $t'_{c1}$, we find a dSC to AF transition driven by $U$. 
For intermediate $t'$ ($t'_{c1}<t'<t'_{c2}$), the AF and dSC phases are separated by a Mott insulator phase (MI), with no AF order (spin liquid phase). 
For $t'>t'_{c2}$, the AF phase disappears, leaving only the dSC and MI phases. 
We find that $t'_{c1}\sim 0.8$ and $t'_{c2}\sim 0.9$.
For $t'>0.25-0.3$, we find a \dxy\ phase at sufficiently low $U$.
The possibility of other magnetic orders, in particular of a $120^\circ$ order at $t'=t$, has not been explored in this work, and therefore we make no claim that the MI phase corresponds to a spin liquid \cite{Kurosaki:2005} everywhere.


\paragraph{Variational Cluster Perturbation Theory.}

V-CPT is an extension of Cluster Perturbation Theory (CPT) \cite{Senechal:2000} based on Potthoff's self-energy-functional approach (SFA)\cite{Potthoff:2003, Dahnken:2004}.
In the SFA one defines a functional $\Omega_{\bf t}[\Sigma]$ of the self-energy that is stationary at the physical self-energy (here ${\bf t}$ stands for the matrix $t_{ij}$ of one-body terms).
This functional can be evaluated exactly for a self-energy $\Sigma'$ that is the exact self-energy of a Hamiltonian $H'$ that differs from the original Hamiltonian only through its one-body terms (i.e., $H$ and $H'$ share the same interaction part).
In V-CPT, the lattice of the model is tiled into identical clusters (Fig.~\ref{fig_dimer}), and $H'$ is the restriction of the dimer model to a finite cluster, to which one adds various Weiss fields that allow for broken symmetries.
The Green function of the Hamiltonian $H'$ is then calculated exactly (by the Lanczos method) and the self-energy functional may then be exactly expressed as\cite{Senechal:2000}
\begin{equation}
\Omega_{\bf t}({\bf t}')=\Omega'\kern-0.1em - \kern-0.1em\int_C \frac{d\omega}{2\pi}\sum_{\mathbf{K}}\ln\det\left(
1 \kern-0.1em + \kern-0.1em (G_0^{-1}\kern-0.2em -G_0'{}^{-1})G'\right)
\end{equation}%
where $G'$ is the exact Green function of $H'$, $G_0$ and $G_0'$ are the non-interacting Green functions of $H$ and $H'$ respectively, and $\Omega'$ is the exact grand potential of $H'$.
A trace over frequencies and wave-vectors of the reduced Brillouin zone is carried, and the Green functions carry discrete indices related to spin and sites within a cluster.
The functional $\Omega_{\bf t}[\Sigma]$ has become a function $\Omega_{\bf t}({\bf t}')$ of the parameters of the cluster Hamiltonian $H'$.
The task is then to find a stationary point of that function, and the exact self-energy of $H'$ at that stationary point, denoted $\Sigma^*$, is adopted as an approximate self-energy for the original model $H$.

\begin{figure}[tbp]
\centerline{\includegraphics[width=8.5cm]{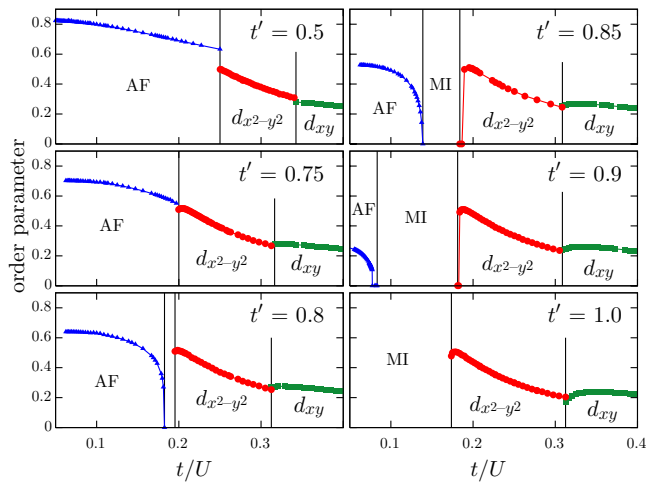}}
\caption{(Color online) Pressure ($t/U$) dependence of the antiferromagnetic (blue triangles), \dxx\ (red circles, scaled by 2) and \dxy (green squares, scaled by 5) order parameter, obtained with V-CPT on $2\times2$ clusters, for various values of $t'/t$.
The vertical lines indicate the transition points, separating the various phases.
}
\label{fig_OP1}
\vglue-5mm
\end{figure}

We insist that the only approximation used is the restriction of the space of self-energies to the set of exact self-energies of a family of cluster Hamiltonians, with the same interaction part as the original Hamiltonian.
In particular, short-range correlation effects are treated exactly through the exact cluster Green function.
In order to study antiferromagnetism and superconductivity, one includes in $H'$ the following ``Weiss terms'':
\begin{eqnarray}
H^{\prime}_{AF}&=&M\sum_{\rv}(-1)^{\sigma} e^{i\Qv\cdot\rv} n_{\rv\sigma}    \\
H^{\prime}_{SC}&=&\sum_{\rv,\rv'}\left(\Delta _{\rv\rv'} c_{\rv\uparrow }c_{\rv'\downarrow}+\mathrm{H.c.}\right)
\end{eqnarray}
where $\Qv=(\pi,\pi)$ is the N\'eel wave vector.
The \dxx\ state is probed by letting $\Delta _{\rv\rv'}=\Delta$ if $\rv'=\rv\pm\xh$ and $\Delta _{\rv\rv'}= -\Delta$ if $\rv'=\rv\pm\yh$.
These Weiss fields allow for the physics of long-range order to seep through the cluster self-energy.
But this is not mean-field theory: the interaction term is never factorized, and the Weiss fields are not the same as the corresponding order parameters. 
The latter can be calculated from the Green function associated with the solution: $G^*{}^{-1} = G_0 - \Sigma^*$.
The density of electrons $n$ is also computed from the Green function $G^*$, as its trace: $n = \Tr\, G^*$.

\paragraph{Gap symmetry.}
The point group of the dimer model is $C_{2v}$, consisting of a rotation of $\pi$ about the $z$ axis, and reflections accross the $\xh+\yh$ and $\xh-\yh$ lines.
This group admits two possible gap symmetries for singlet superconductivity: the \dxx\ state defined above ($A_2$ representation), and the isotropic state ($s$-wave, or $ A_1$ representation).
See Ref.~\onlinecite{Powell:2006} for a more detailed group-theoretical analysis.
We also included in our study a SC Weiss fields belonging to the $A_1$ representation:
\begin{eqnarray}
\Delta_{\rv,\rv+\xh+\yh} = \Delta_{\rv,\rv-\xh-\yh} &=& \phantom{-}\Delta_1 \\
\Delta_{\rv,\rv+\xh-\yh} = \Delta_{\rv,\rv-\xh+\yh} &=& -\Delta_2 
\end{eqnarray}
(note that the on-site Coulomb repulsion excludes the possibility $\Delta_{\rv,\rv}\ne0$).
If $\Delta_1=\Delta_2$, these two possibilities form together what is customarily called the \dxy\ state, but the relative values of $\Delta_1$ and $\Delta_2$ are not constrained by the symmetry of the model.
We will nevertheless refer to this state as \dxy.
Even though this state belongs to the $A_1$ representation, the corresponding gap function will display nodes (roughly along the $x$ and $y$ axes), even though these nodes may not be robust -- e.g. with respect to impurities -- on group-theoretical grounds \cite{Powell:2006}.
Recent STM observations\cite{Ichimura:2006} are interpreted as favoring this state.

In this work, five variational parameters were used: the Weiss fields $M$, $\Delta$, $\Delta_1$ and $\Delta_2$, as well as the chemical potential $\mu'$ on the cluster, the latter in order to ensure thermodynamic consistency\cite{Aichhorn:2005}.
The cluster chemical potential $\mu'$, which is part of $H'$, must not be confused with the actual chemical potential $\mu$, which is part of $H$ and controls the density of electrons on the lattice.
The Nambu formalism was used in order to treat anomalous averages.
Time-reversal violating states such as $d+is$ (e.g. a complex mixture of the $A_1$ and $A_2$ representations) were not investigated, because of our use of real-number quantum mechanics in numerical computations.
\begin{figure}[tbp]
\centerline{\includegraphics[width=8.5cm]{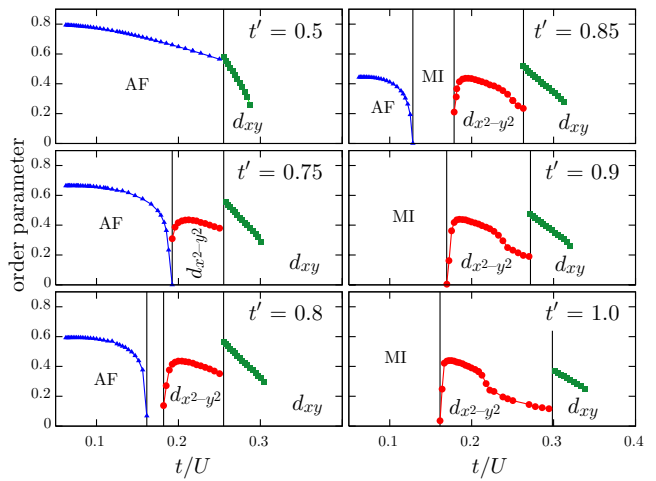}}
\caption{(Color online) Same as Fig.~\protect\ref{fig_OP1}, this time for $2\times4$ clusters.
}
\label{fig_OP2}
\vglue-5mm
\end{figure}

\paragraph{Results.}
Fig.~\ref{fig_OP1} shows the order parameters as a function of pressure $t/U$, for different values of $t'$, using a 4-site cluster ($2\times2$).
Fig.~\ref{fig_OP2} shows the same type of data obtained on an 8-site cluster ($2\times4$). 
The blue triangles were obtained by treating only $M$ and $\mu'$ as variational parameters, i.e. by forbidding a SC solution.
The red circles were obtained by treating only $\Delta$ and $\mu'$ as variational parameters, i.e. by forbidding an AF solution.
The green squares were obtained by treating $\Delta_1$, $\Delta_2$ and $\mu'$ as variational parameters.
No mixed solutions were found by performing a 3-parameter variation of $M$, $\Delta$ and $\mu'$.
Normal state solutions, where the only variational parameter is $\mu'$, were also obtained.
For each value of $U$ and $t'$, the lattice chemical potential $\mu$ was adjusted in order for the electron density $n$ to be unity (half-filling) within 0.1\%.
Then a comparison of the ground state energy per site $\Omega+\mu n$ was made in order to decide which of the four solutions (AF, \dxx, \dxy\ or normal) was preferred (only the lowest-energy solutions are shown here).
On Figs \ref{fig_OP1} and \ref{fig_OP2}, vertical lines separate the different phases thus determined, with the SC phases having the lowest energy at high pressure, the AF phase at low pressure, with an intercalated normal phase -- in fact a paramagnetic Mott insulator -- appearing for $t'$ greater than some critical value $t'_{c1}$ that is close to 0.8.
The AF phase disappears altogether for $t'$ greater than some critical value $t'_{c2}$ of about 0.9.

The transitions between the different ordered states (AF, \dxx\ and \dxy) are of the first order, whereas the transitions from AF to normal look continuous.
The results obtained on the 4-site and 8-site clusters are in overall agreement on the existence of all four phases and their relative location.
The critical value $U_c(t')$ for the SC to AF/MI transition increases slightly with $t'$, but stays between 5 and 6 for both cluster sizes, at least for $t'\geq 0.75$.
The maximum magnitude of the \dxx\ order parameter does not vary much with $t'$.
By contrast, the critical $U$ for the MI to AF transition  increases sharply with $t'$, as the saturation value of the AF order parameter drops.
Differences between the two cluster sizes lie mostly in the extent of the \dxy\ phase, which is more important in the 8-site than on the 4-site cluster.
This is reflected in the ``foggy'' boundary between the two SC phases that we display on the schematic phase diagram of Fig.~\ref{fig_dimer}.
Note that the cluster sizes used are too small to perform any kind of finite-size analysis, and we cannot say that the results are converged.

The \dxy\ phase only appears at sufficiently high $t'$ (it was not found at $t'=0.25$) and low $U$: it seems favored by the dispersion, but at intermediate coupling the tide turns towards a \dxx\ order.
In this V-CPT calculation, there is no evidence of a lower critical $U$: the SC phase appears to extend all the way to $U=0$ and no normal metallic phase is found.

\begin{figure}[tbp]
\centerline{\includegraphics[width=7cm]{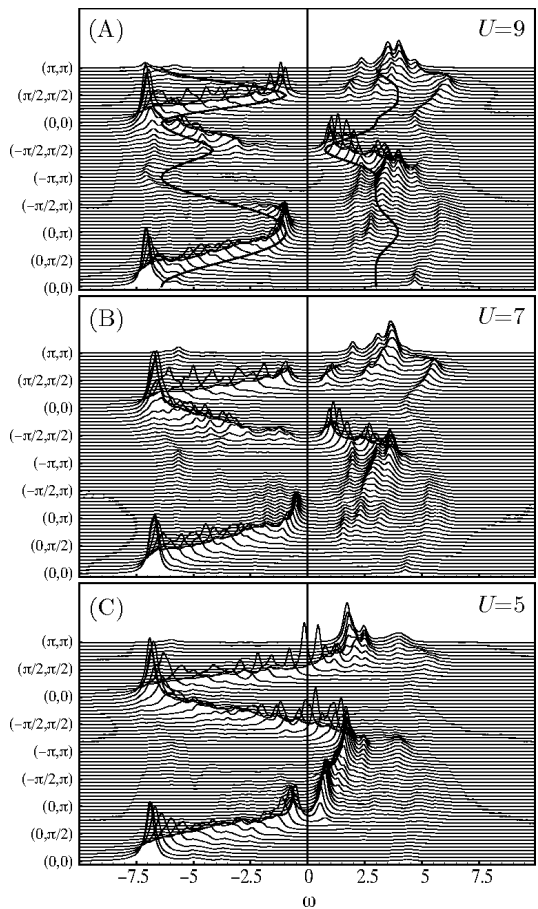}}
\caption{
Energy distribution curves (i.e. spectral function) for three sample solutions with $t'=0.8t$.
(A): an antiferromagnetic solution at $U=9t$.
The SDW dispersion curves for the same values of the parameters ($U$, $\mu$, $t'$ and $t$) and a mean field $h=0.4t$ are also drawn.
(B): the same, for a Mott insulator solution ($U=7t$).
What looked like the AF gap in (A) is now a Mott gap.
(C): the same, for a \dxx\ solution ($U=5t$). 
Notice the $d$-wave gap maximum along the $(0,0)-(0,\pi)$ segment.
In all cases, a Lorentzian broadening of $0.15t$ has been used.
}
\label{fig_edc}
\vglue-4mm
\end{figure}

Fig.~\ref{fig_edc} shows the spectral function for sample solutions for three different values of $U$ corresponding to (A) the AF phase, (B) the MI phase and (C) the \dxx\ phase at $t'=0.8t$, on an 8-site cluster.
In (A) the mean-field AF dispersion is also plotted (full lines), with a mean-field $h=0.4t$, a value chosen so as to best match the V-CPT dispersions (the actual V-CPT order parameter in that case is $\L M\R = 0.497$).
Note that the mean-field dispersion roughly matches the V-CPT spectral function, at least where the weight is important.
In (B), the spectral function displays a Mott gap across all wave-vectors.
It is very similar to the AF spectral function (A), except that it does not curve back to follow the AF mean-field dispersion near $(0,\pi)$ and $(\pi/2,\pi/2)$.
In (C), we notice the \dxx\ gap, maximum along the $(0,0)-(0,\pi)$ segment, and vanishing along the $(-\pi,\pi)-(0,0)$ segment.
Except for that dSC gap, the spectral function is that of a metal, i.e. there is no other gap in the spectrum.
Unfortunately, photoemission (ARPES) experiments, which would offer a probe on the negative-frequency part of the spectral function, have not, to our knowledge, been performed on these compounds.

To conclude, the dimer model of layered organic conductors captures the essential features of this family of compounds: AF and SC phases separated by a first order transition, except at higher frustration levels where a MI phase (possibly spin liquid) appears.
We found that the superconducting gap symmetry changes from the $A_1$ (i.e., \dxy) to the $A_2$ (i.e., \dxx) representation of the point group as $U$ increases, which leads to the possibility of a close interplay between those two states as a function of pressure in these compounds.
Of course an exciting development would be the synthesis of a new member of the family displaying N\'eel order at ambient pressure, with a value of $t'$ in the 0.8-0.9 range.
Then this compound would be predicted to go through a spin liquid, and then through a dSC phase upon applying pressure.


\begin{acknowledgments}
We are indebted to A.-M. Tremblay, C.~Bourbonnais, B.~Kyung and D. Fournier for useful discussions.
This work was supported by NSERC (Canada).
Computations were performed on the Dell clusters of the R\'eseau Qu\'eb\'ecois de Calcul de haute performance (RQCHP).
\end{acknowledgments}


\end{document}